\documentclass[letter]{aa}  

\usepackage{graphicx}
\usepackage{txfonts}
\linenumbers

\begin{document}

   \title{Evidence for inflows and outflows in the nearby black hole transient Swift J1727.8$-$162}
   \titlerunning{Inflows and outflows in the black hole transient Swift J1727.8$-$162}
   \author{D. Mata S\'anchez\inst{1,2}
          \and
          T. Mu\~noz-Darias\inst{1,2}          
          \and
          M. Armas Padilla\inst{1,2}
          \and
          J. Casares\inst{1,2}
          \and
          M. A. P. Torres\inst{1,2}
          }

   \institute{Instituto de Astrof\'isica de Canarias, E-38205 La Laguna, Tenerife, Spain\\
              \email{matasanchez.astronomy@gmail.com, dmata@iac.es}
         \and
             Departamento de Astrof\'isica, Univ. de La Laguna, E-38206 La Laguna, Tenerife, Spain\\
             }

   \date{Received 27 November 2023. Accepted 8 January 2024}

  \abstract
  {We present 20 epochs of optical spectroscopy obtained with the GTC-10.4m telescope across the bright discovery outburst of the black hole candidate Swift J1727.8$-$162. The spectra cover the main accretion states and are characterised by the presence of hydrogen and helium emission lines, commonly observed in these objects. They show complex profiles, including double peaks, but also blue-shifted absorptions (with blue-edge velocities of $1150 \, {\rm km\, s^{-1}}$), broad emission wings, and flat-top profiles, which are the usual signatures of accretion disc winds. Moreover, red-shifted absorptions accompanied by blue emission excesses suggest the presence of inflows in at least two epochs, although a disc origin cannot be ruled out. Using pre-outburst imaging from Pan-STARRS,  we identify a candidate quiescent optical counterpart with a magnitude of $g \sim 20.8$. This implies an outburst optical amplitude of $\Delta V \sim 7.7$, supporting an estimated orbital period of $\sim 7.6\, {\rm h}$, which favours an early K-type companion star. Employing various empirical methods, we derive a distance to the source of $d = 2.7 \pm 0.3 \, {\rm kpc}$, corresponding to a Galactic plane elevation of $z = 0.48 \pm 0.05 \, {\rm kpc}$. Based on these findings, we propose that Swift J1727.8$-$162 is a nearby black hole X-ray transient that exhibited complex signatures of optical inflows and outflows throughout its discovery outburst.}

   \maketitle
%

\section{Introduction}

Swift~J1727.8$-$162 (hereafter J1727) was originally identified as a gamma-ray burst (GRB 230824A; \citealt{Lipunov2023,Page2023}) by \textit{Swift} on August 24,  2023 (MJD 60180.5831; all times will be referred to this date). The duration and X-ray colour of the event, however, led to a new improved classification as a low-mass X-ray binary, whose outburst evolution was immediately followed by the  \textit{MAXI/GSC} \citep{Negoro2023,Nakajima2023} and \textit{NICER} \citep{Oconnor2023} observatories. The source evolved through the X-ray hard and soft states  (see e.g. \citealt{Bollemeijer2023,Miller-Jones2023}), following the canonical behaviour for black hole transients (BHTs; e.g. \citealt{Done2007,Belloni2011}). The brightness of the event (see e.g. \citealt{Palmer2023}) prompted multiwavelength campaigns with ground-based monitoring. Optical follow-up revealed a very bright $r\sim 12.7$ counterpart \citep{Wang2023,Baglio2023,Alabarta2023}, which subsequently started a slow decline over the outburst duration. This enabled spectroscopy from the early stages of the outburst \citep{CastroTirado2023}, revealing typical Doppler-broadened emission lines of H and He species. \citet{MataSanchez2023c} further extended the identification to other metallic transitions while reporting on the detection of blue-shifted absorption features in a number of lines, a hallmark of cold disc outflows. 

In this letter we confirm the detection of outflows through the analysis of multi-epoch optical spectroscopy obtained during the first 62 days of the outburst and across all the classical accretion states. Additionally, we present a pre-outburst optical counterpart and provide constraints on some fundamental parameters, such as the orbital period and the distance to the source.

\begin{table*}
\label{tab:obslog}
\caption{Journal of observations.}
\begin{tabular}{ccccccccccc}
Epoch & Date & TST  & Grism+ & R &$$\#$$ & $\rm {T_{exp}}$ & Seeing & Airmass &  X-ray state \\
 &(dd/mm) & (d)& slit width (\arcsec) & $\lambda /\Delta \lambda$ &  & (s) &(\arcsec) &  \\
\hline \\
E1& 26/08 & 2.293 & R1000B+0.8 & 1180 & 4 & 150 & 0.9& 1.44& Hard\\
E2& 27/08 & 3.312 & R1000B+0.8 & 1180 & 18 & 150 & 1.2& 1.51& Hard\\
E3& 30/08 & 6.306 & R1000B+0.6 & 1230 & 3 & 150 & 1.2& 1.51& Hard\\
E4& 02/09 & 9.312 & R2500R/R2000B+0.8 & 2830/2420  & 1,1 & 250 & 1.3& 1.56& HIMS\\
E5& 04/09 & 11.303 & R2500I/R/U+0.8 & 2800/2940/1900 & 1/1/1 & 250 & 1.0 & 1.54& HIMS \\
E6& 05/09 & 12.298 & R2500R/V/U+0.8 & 2970/3220/1900 & 1/1/1 & 250 & 0.8 & 1.54& HIMS\\
E7& 06/09 & 13.347 & R2500R/V/U+0.8 & 3060/3190/1900 & 1/1/1 & 250 & 0.9& 1.94& HIMS\\
E8& 09/09 & 16.269 & R2500R/V/U+0.8 & 2880/3230/1900 & 1/1/1 & 250 & 1.1& 1.47& HIMS\\
E9& 10/09 & 17.299 & R2500R/V/U+0.8 & 2970/3190/1900 & 1/1/1 & 250 & 0.9& 1.61& HIMS\\
E10& 21/09 & 28.275 & R2500R/R2000B+0.8 & 2970/2520 & 1/1 & 250 & 1.3& 1.65& HIMS\\
E11& 22/09 & 29.307 & R2500R/R2000B+0.8 & 2970/2520 & 1/1 & 250 & 1.4& 1.98& HIMS\\
E12& 24/09 & 31.276 & R2500R/R2000B+0.8 & 2940/2480 & 1/1 & 250 & 1.3& 1.71& HIMS\\
E13& 27/09 & 34.287 & R2500R/R2000B+0.8 & 3030/2540 & 1/1 & 250 & 1.8& 1.82& HIMS\\
E14& 01/10 & 38.286 & R2500R/R2000B+0.8 & 3060/2560 & 1/1 & 250 & 0.9& 2.03& HIMS\\
E15& 05/10 & 42.276 & R2500R/R2000B+0.8 & 3000/2600 & 1/1 & 250 & 1.0& 2.06& SIMS\\
E16& 06/10 & 43.251 & R2500R/R2000B+0.8 & 2970/2560 & 1/1 & 250 & 1.1& 1.83& SIMS\\
E17& 09/10 & 46.252 & R2500R/R2000B+0.8 & 2970/2650 & 1/1 & 250 & 0.9& 1.89& HIMS\\
E18& 11/10 & 48.263 & R2500R/R2000B+0.8 & 3030/2560 & 1/1 & 250 & 1.0& 2.13& HIMS\\
E19& 13/10 & 50.236 & R2500R/R2000B+0.8 & 2860/2540 & 1/1 & 250 & 1.3& 1.83& HIMS\\
E20& 24/10 & 61.228 & R2500R/R2000B+0.8 & 2830/2460 & 1/1 & 250 & 2.3& 2.13& Soft\\
\end{tabular}
\tablefoot{TST stands for `time since trigger', defined as the difference between the mid-exposure \rm{MJD} of each spectroscopic epoch and the \textit{Swift} discovery trigger at $\rm{MJD}\, 60180.5831$ (2023-08-24 13:59:44 UTC; \citealt{Page2023}). We employed a $0.6\arcsec$ or $0.8\arcsec$ slit width and a number of grisms, each covering a different wavelength range: R1000B ($3600 - 7800\, {\rm \AA}$), R2000B ($3940 - 5680\, {\rm \AA}$), R2500U ($3420 - 4600\, {\rm \AA}$), R2500V ($4420 - 6050\, {\rm \AA}$), R2500R ($5560 - 7670\, {\rm \AA}$), and R2500I ($7300 - 10140\, {\rm \AA}$). The spectral resolution was measured from lines in the comparison arc lamps. The number of exposures per epoch ($\#$) and their individual exposure times ($\rm {T_{exp}}$) are also reported. }
\end{table*}

\section{Observations} \label{sec:observations}

We obtained 20 spectroscopic epochs with the \mbox{10.4m} Gran Telescopio Canarias (GTC) at the Roque de los Muchachos Observatory (La Palma, Spain), equipped with the Optical System for Imaging and low-Intermediate-Resolution Integrated Spectroscopy (OSIRIS, \citealt{Cepa2000}). The observing set-up varied across the different epochs, as detailed in Table \ref{tab:obslog}. The spectra were reduced, extracted, and wavelength-calibrated in \textsc{pyraf} and \textsc{molly}. We employed sky emission lines to correct from flexure effects on the wavelength calibration, finding sub-pixel drifts.

\section{Analysis and results} \label{sec:analysis}

\subsection{Hardness--intensity diagram} \label{sec:hid}

\begin{figure}
   \centering
   \includegraphics[keepaspectratio, trim=0cm 0cm -0.5cm 0cm, clip=true, width=0.5\textwidth]{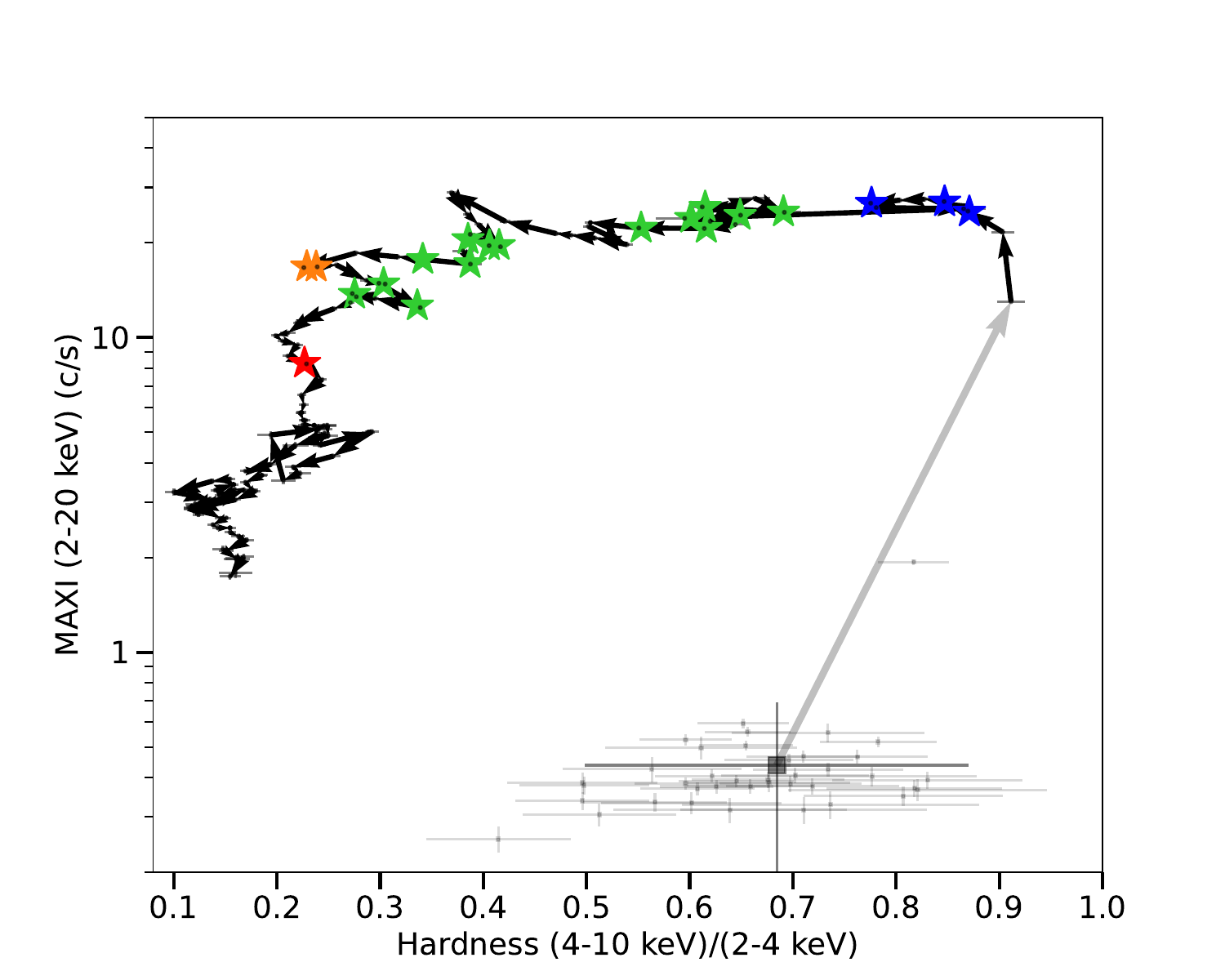}
   \caption{HID constructed from \textit{MAXI} data. Hardness corresponds to the ratio of the count rates in the 4-10 keV to the  2-4 keV bands, while the intensity corresponds to the count rate in the 2-20 keV band. The black line depicts the outburst evolution with the grey square representing a binning of $\sim 50$ days of data prior to the X-ray trigger (shown as transparent points in the background). The stars mark the timestamps of our optical observations, coloured according to the X-ray state (see Table \ref{tab:obslog}): hard state (blue), HIMS (green), SIMS (orange), and soft state (red).}
        \label{fig:hid}
    \end{figure}

In order to put our observations in context with the outburst evolution, we built a  hardness--intensity diagram (HID, \citealt{Homan2001}) employing daily averaged fluxes from \textit{MAXI} (see Fig. \ref{fig:hid}). The HID follows the canonical q-shaped hysteresis pattern observed in BHTs (e.g. \citealt{Fender2012}), characterised by lower hardness values than their neutron star counterparts (e.g. \citealt{Munoz-Darias2014}). The X-ray state of each GTC epoch is listed in Table \ref{tab:obslog}. Inspection of the HID reveals that there were X-ray detections prior to the outburst trigger. During the initial outburst rise in the hard state, quasi-periodic oscillations (QPOs) were detected evolving from 0.4 Hz to 1.4 Hz \citep{Palmer2023,Draghis2023,Katoch2023,Debnath2023,Katoch2023b}. The source entered the hard-intermediate state (HIMS) in day 9, as QPOs developed at higher frequencies ($4-6\, {\rm Hz}$, \citealt{Bollemeijer2023,Mereminskiy2023}). Transition to the soft-intermediate state (SIMS) occurred a month later, based on the disappearance of the QPO and a drop in the aperiodic X-ray variability \citep{Bollemeijer2023B,Belloni2005}. On day 52 a transition to the soft state was reported \citep{Trushkin2023}, where the system gradually decayed in luminosity. At the time of writing this letter, the outburst is still ongoing.

\subsection{Evolution of the optical spectra}

\begin{figure*}
\centering
\includegraphics[keepaspectratio, trim=2.5cm 0cm 2.5cm 3cm, clip=true, width=0.7\textwidth]{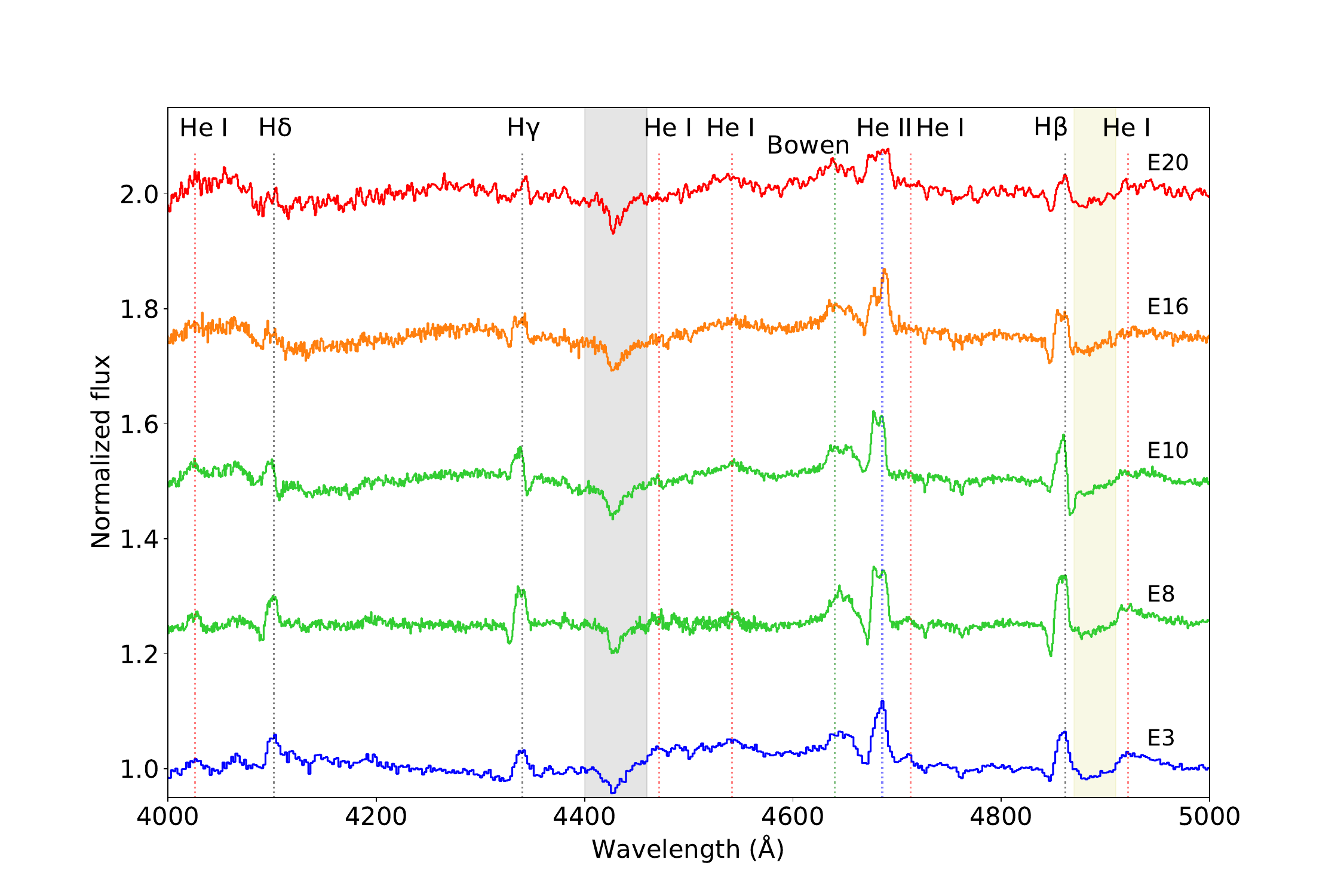}\includegraphics[keepaspectratio,  trim=0cm 0cm 0.5cm 3cm, clip=true, width=0.26\textwidth]{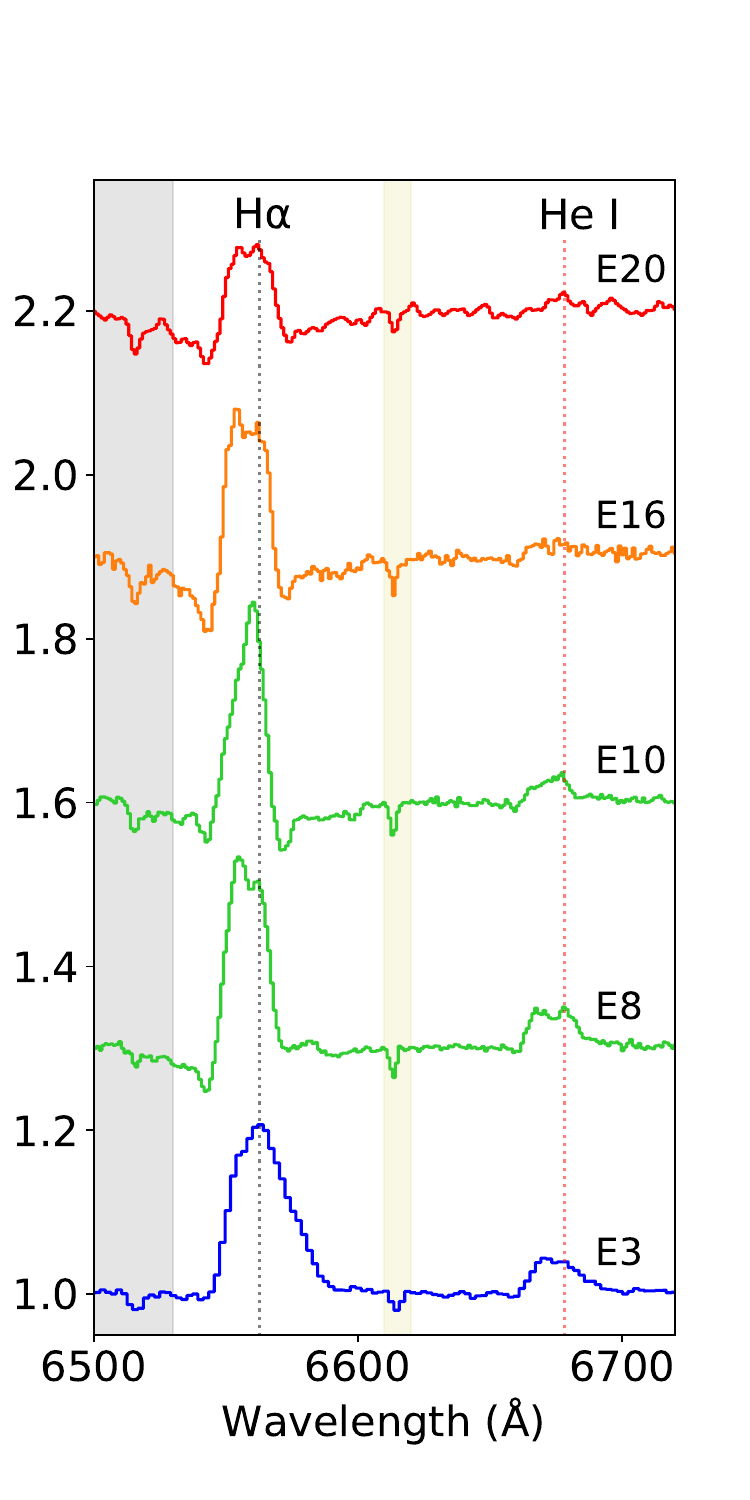}
   \caption{Normalised spectrum of five relevant spectroscopic epochs. They are offset vertically by 0.25 (0.3 in the right panel) for visualisation purposes; the  colours correspond to their X-ray state, as defined in Fig. \ref{fig:hid}. The identified transitions in this wavelength range are marked with dotted lines:  black for H, red for \ion{He}{i}, blue for \ion{He}{ii}, and green for other metallic lines (Bowen blend at $\sim 4640\,\rm{\AA}$). Telluric bands and interstellar features are shown as  grey and yellow shadowed regions, respectively. The E20 spectrum was smoothed through convolution with a Gaussian for visualisation purposes.}
        \label{fig:norm}%
    \end{figure*}

Five different epochs, representative of the collected data during the four different X-ray states, are shown in Fig. \ref{fig:norm}. The spectrum of J1727 is characterised by the presence of emission lines corresponding to H Balmer and Paschen series, \ion{He}{i} (e.g. $5875.618\,\rm{\AA}$, $6678.149\,\rm{\AA}$) and \ion{He}{ii} ($4685.750\,\rm{\AA}$, $5411.551\,\rm{\AA}$) transitions. The Bowen blend and emission from the \ion{Fe}{ii} forest ($\sim 5284\,\rm{\AA}$) are also detected. We searched for short-term variability in our extended E2 observations (a total of 18 spectra), but could not see evidence of any within timescales of a few minutes to an hour. While this does not preclude stronger variability from occurring at a different state of the outburst, we   base our study on the average spectral properties obtained for each epoch.

An overall view of the evolution of the emission lines reveals predominantly single-peaked lines during the bright hard state (E1 to E3), while at later epochs (HIMS) they develop a double-peaked structure, first on the \ion{He}{i} and \ion{He}{ii} transitions, (starting on E4) and later over the H series (e.g. E16). We performed single and two-Gaussian fits to each $\rm{H{\alpha}}$, $\rm{H{\gamma}}$, and $\rm{H{\delta}}$ line, as the remaining transitions are too weak (even undetectable at certain epochs, see e.g. \ion{He}{i}-$6678\,\rm{\AA}$) or contaminated by nearby features (e.g. \ion{He}{i}-$5876\,\rm{\AA}$, $\rm{H{\beta}}$). In spite of the presence of variable asymmetric components distorting the profiles (described below), the single-Gaussian fit yields an average full width at half maximum of $\rm{FWHM}= 750\pm 120\, {\rm km\, s^{-1}}$. The two-Gaussian fit leads to a consistent centroid velocity across all epochs of $-170 \pm 50\, {\rm km\, s^{-1}}$. Given the multi-epoch stability of the line centroid, we   adopt it as the tentative radial systemic velocity of J1727. We   refer any line profile to this value  (i.e. in the binary rest frame). We note that a Gaussian fit to \ion{He}{ii}-$4686\,\rm{\AA}$ yields a consistent centroid, although the FWHM increases monotonically as the outburst progresses, reaching $\rm{FWHM}= 1500\pm 300\, {\rm km\, s^{-1}}$ in the soft state observation (E20, see Fig. \ref{fig:norm}).

\begin{figure*}
\centering
\includegraphics[keepaspectratio, trim=0cm 0cm 0cm 0cm, clip=true, width=\textwidth]{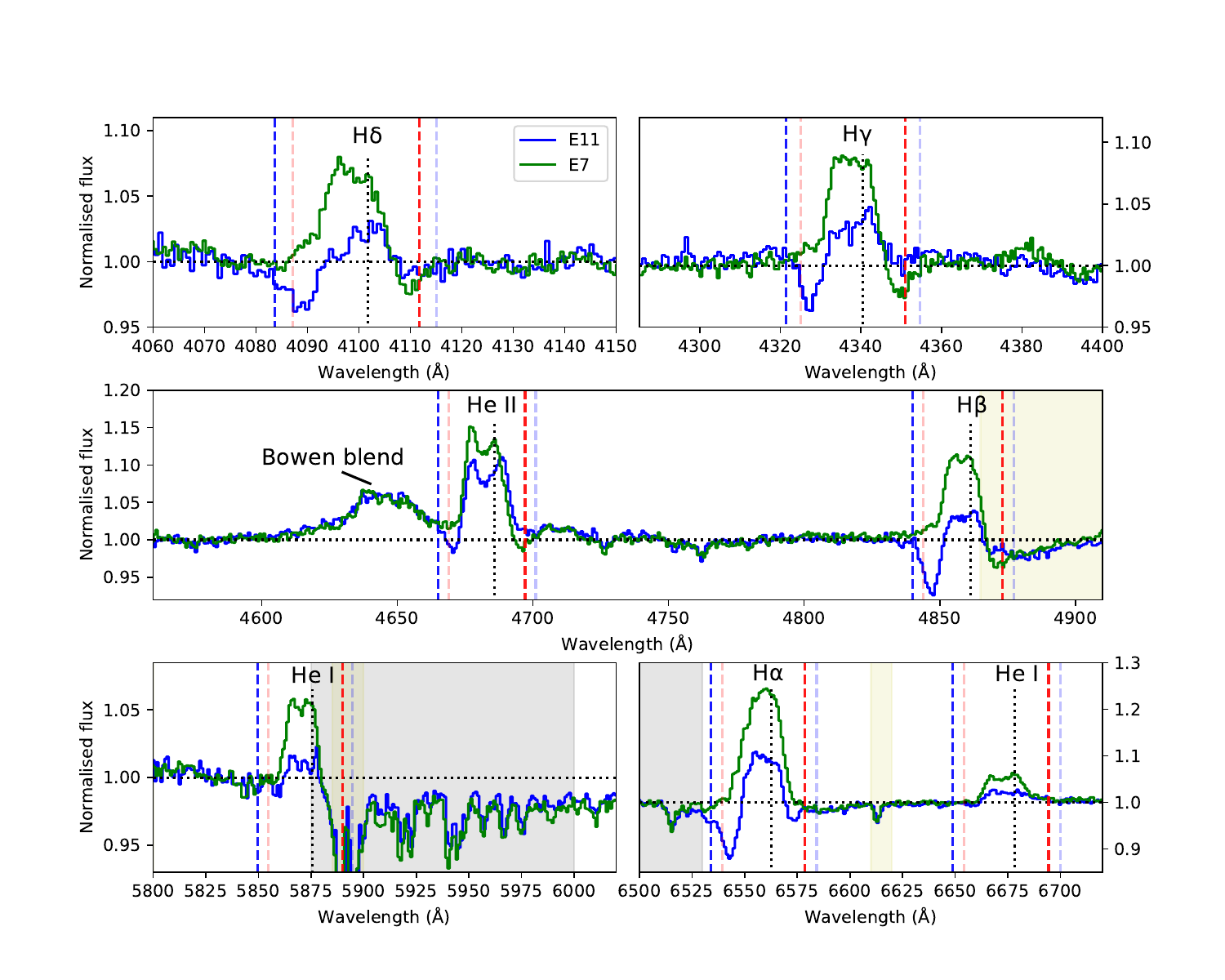}
   \caption{Normalised spectrum of epochs E7 (green) and E11 (blue). They are representatives of the blue-shifted and red-shifted absorption features detected during the outburst. The identified transitions in this wavelength range are marked with black dashed lines (in the laboratory rest frame), while the velocities of $\pm 900\, {\rm km\, s^{-1}}$ and $\pm 1150\, {\rm km\, s^{-1}}$ (referred to the rest wavelength of each line after accounting for the systemic velocity) are marked with red and blue dashed lines, respectively. Telluric bands and interstellar features are shown as  grey and yellow shadowed regions, respectively.}
        \label{fig:lines}%
    \end{figure*}

$\rm{H{\alpha}}$ and \ion{He}{i}-$5876\,\rm{\AA}$ are usually the best lines to search for outflows. In the case of J1727, the presence of nearby and strong telluric features hampers a clean detection. $\rm{H{\beta}}$ is similarly affected by the presence of a diffuse interstellar band (DIB) that contaminates the red wing of the line. For this reason we initially focus our analysis on $\rm{H{\gamma}}$ and $\rm{H{\delta}}$. We find two remarkable patterns on the profiles (see Figs. \ref{fig:norm} and \ref{fig:lines} for examples of the epochs cited below):

\begin{itemize}
    \item Blue-shifted absorptions: They consistently appear in the Balmer series across most epochs. \ion{He}{i} transitions do not show such clear patterns, but \ion{He}{ii}-$4686\,\rm{\AA}$ does, dipping below the continuum (see e.g. E11). The absorption depth is variable, being at its deepest in E8 and E11, reaching $7\%$ below the continuum level for $\rm{H{\beta}}$ ($4\%$ for $\rm{H{\gamma}}$). They show blue-edge velocities of $\sim -1150 \, {\rm km\, s^{-1}}$, which remain consistent across all epochs. 
    
    \item Red-shifted absorptions: A number of epochs exhibit red-shifted absorptions (E4, E7, E9, E10, E19). They meet the continuum at $\sim 900 \, {\rm km\, s^{-1}}$,    a velocity $\sim 20\%$ lower than that of  their blue-shifted counterparts, and exhibit slightly shallower depths. The deepest absorption is found in epoch E10 (a day before the deepest blue-shifted absorption was observed) being $2\%$ below the continuum level for $\rm{H{\gamma}}$. These red-shifted absorptions can be accompanied by blue-shifted absorptions (E10), broad blue wings (E4, E7), skewed (E9, E10), or even flat-top (E7) emission line profiles. We note that even the high excitation \ion{He}{ii}-$4686\,\rm{\AA}$ line exhibited these red-shifted absorptions in E4 and E7.

\end{itemize}

\section{Discussion} \label{sec:discussion}

\subsection{Optical accretion disc winds}

We detect blue-shifted absorptions (E11, Fig. \ref{fig:lines}) sometimes accompanied by a red tail excess (E3, Fig. \ref{fig:norm}), matching the prescription for P-Cygni profiles (e.g. \citealt{Munoz-Darias2016,SanchezSierras2023}). They persist across the outburst, predominantly in the H Balmer series, with stable terminal velocities ($\sim -1150 \, {\rm km\, s^{-1}}$) consistent with previous studies in other BHTs (see \citealt{Panizo-Espinar2022}). In addition, these systems can also show broad emission wings that meet the continuum at the same velocity as the P-Cygni profiles. These features are thought to be associated with optically thin ejecta, as well as ultimately responsible for the so-called nebular phase (e.g. \citealt{MataSanchez2018}). Flat top (E7, see Fig. \ref{fig:lines}) and skewed (E9, E10, see Fig. \ref{fig:norm}) profiles, both typically associated with the presence of outflows, are also identified. All this phenomenology (see Table \ref{tab:outflows} for a compilation) suggests the continuous presence of outflows signatures during the hard state, HIMS, and SIMS, as shown by other BHTs in outburst (e.g. \citealt{Munoz-Darias2019,Panizo-Espinar2022}).

The analysis of the J1727 spectra is complex due to the presence of contaminants and the multi-component nature of the line profiles. While the effect of DIBs and telluric absorptions can be mitigated by analysing clean wavelength regions, the detection of red-shifted absorptions at certain epochs poses a challenge. We propose two alternative scenarios to explain them. On the one hand, the simultaneous presence of a weaker blue-shifted absorption component during epochs with red-shifted absorptions (e.g. Balmer series in E10) suggests that the emission line might be embedded into a broad absorption component. These have been observed in H and \ion{He}{i} transitions in other BHTs (e.g. GRO J1655$-$40; \citealt{Soria2000}). Their origin is still debated, as predictions from the most commonly accepted model \citep{Dubus2001} have been challenged (see e.g. \citealt{MataSanchez2022} and references therein). On the other hand, narrow red-shifted absorptions have been previously reported in the optical spectra of the BHT GRS 1716$-$249 \citep{Cuneo2020b}, and described as inverse P-Cygni profiles (i.e. inflow signatures). In our dataset, the line profiles of $\rm{H{\gamma}}$ and $\rm{H{\delta}}$ in E4 and E7 (see Fig. \ref{fig:lines}) are the most promising candidates supporting this scenario, as they are concurrent with extended blue emission wings. To shed some light on this matter, we searched for outflows and broad absorption features with the machine learning classifier ATM \citep{MataSanchez2023b} adapted to the $\rm{H{\gamma}}$ and $\rm{H{\delta}}$ lines (which are free from nearby contaminants). The resulting classifications (collected in Table \ref{tab:outflows}) support the detection of outflow features (mainly as blue-shifted absorptions) in several epochs, while a number of profiles are classified as embedded in a broad absorption component (including some with narrow red-shifted absorption features; e.g. E7, E19). A Monte Carlo approach reveals, for spectra where the classifier struggles the most, that the broad absorption class is the second most likely classification (typically $\sim 30-40\%$ association probability). This suggests that a broad absorption component might be present during most of the outburst, but only dominates the profile at certain epochs. The absence of a dedicated inflow class in ATM precludes further inspection.

Another puzzling result is the detection of possible outflow features in \ion{He}{ii}-$4686\,\rm{\AA}$, including blue-shifted absorptions dipping below the continuum (E11, Fig. \ref{fig:lines}) and skewed profiles (E3, see Fig. \ref{fig:norm}). This line behaves as the rest of the simultaneously observed transitions in these epochs, both in terms of depth and blue-edge velocity, which leads us to conclude that it traces the same phenomena. This would be one of the few detections of outflows in a high-ionisation optical transition, with Swift J1357.2$-$0933 being the only precedent \citep{Charles2019,JimenezIbarra2019b}.

The only observation in our sample during the soft state is E20 (Fig. \ref{fig:norm}). By then, $\rm{H{\gamma}}$ exhibits a skewed profile (where the blue peak is absorbed, compared to other symmetric double-peaked transitions). There is a blue-shifted absorption in $\rm{H{\beta}}$ reaching $4\%$ below the continuum (comparable to E16, see Fig. \ref{fig:norm}). However, the contaminants in the red wing of $\rm{H{\beta}}$, as well as the lack of a similar feature in other lines, preclude a definitive confirmation. An unambiguous detection of optical outflows during the soft state of a BHT remains elusive (but see \citealt{Ponti2012,SanchezSierras2020,Munoz-Darias2022,Parra2023} for wind detections at infrared and X-ray wavelengths).

\subsection{Quiescent counterpart and orbital period}

\begin{figure}
   \centering
   \includegraphics[keepaspectratio, trim=3.5cm 1cm 3cm 1cm, clip=true, width=0.5\textwidth]{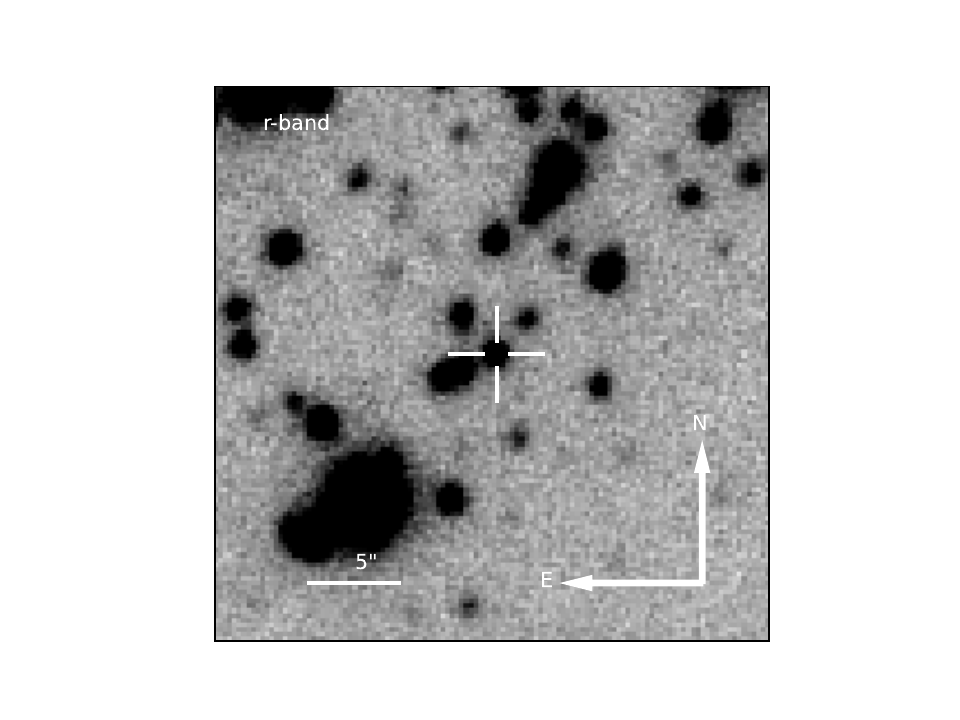}
   \caption{Pre-outburst quiescent counterpart of J1727 from PS1 r-band images. The position of the quiescent counterpart is marked with a white cross.}
        \label{fig:fc}
\end{figure}

Using the radio coordinates [R.A. (J2000) 17:27:43.31 $\pm$ 0.04s, Dec. (J2000): -16:12:19.23 $\pm$ 0.02$\arcsec$] given by \citet{MillerJones2023},  we searched for the quiescent counterpart in all-sky public photometric catalogues. We found a single candidate in both Pan-STARRS DR1 (PS1, \citealt{Chambers2016}; ID 88552619307433951) and in \textit{Gaia} DR3 (\citealt{Gaia2016,Gaia2023}; ID 4136619720186379264). The Gaia and radio positions, after correcting for the proper motion, are consistent within $2.5\sigma$ ($\sim 0.07\, ''$; see Fig. \ref{fig:fc}). We consider this matching object the quiescent counterpart. In the calculations that follow we  use the mean PS1 PSF magnitudes: $g=20.82\pm0.05$, $r=19.95\pm0.04$, $i=19.41\pm0.02$, $z=19.06\pm0.02$.

Determination of the orbital period during the outburst usually relies on either the detection of eclipses, dips (both limited to high-inclination orbits), or photometric modulations. Neither have been reported so far for J1727. We use the correlation in \citet{Shahbaz1998} connecting this binary parameter with the V-band outburst amplitude:
$$\Delta V= 14.36-7.63 \, \log{(P_{\rm orb}(\rm{h}))}.$$

To calculate $\Delta V$ we take the outburst peak V-band magnitude $V_{\rm outburst}=12.66\pm 0.01$ (\citealt{Alabarta2023}), measured on day 4.8 (simultaneous to the peak at X-rays). We established a quiescent V-band magnitude of $V_{\rm quiescence}=20.31\pm0.06$ applying photometric system transformations \citep{Jordi2006,Tonry2012} to the PS1 pre-outburst photometry. This leads to $\Delta V=7.65\pm 0.06$ and $P_{\rm orb}=7.57\pm0.14\, {\rm h}$. Following \citet{Faulkner1972}, the combination of such $P_{\rm orb}$ with the assumption of a Roche-Lobe filling dwarf companion star (with mean density as per \citealt{Drilling2000}) imply an early K spectral type. This result should be taken with some caution as different systematics might be at play (see e.g. \citealt{Lopez2019}).

\subsection{The distance to J1727}

No radio parallax is known to date for J1727, while the parallax from Gaia DR3 ($\pi = 0.03 \pm 0.39\, \rm{m.a.s.}$) results in an unconstraining posterior distribution (see e.g. XTE J1118+480 in \citealt{Gandhi2019}). We thus here employ different methods to constrain its distance and reddening.

\citet{Casares2018a} presents an empirical correlation between the absolute r-band magnitude for BHTs in quiescence and $P_{\rm orb}$:
$$M_{\rm r}=(4.64\pm 0.10) - (3.69\pm 0.16)\, \log{(P_{\rm orb}(\rm{d})).}$$
Adopting $P_{\rm orb}= 7.57\pm 0.14\,{\rm h}$ we obtain $M_{\rm r}=6.50\pm 0.13$. $M_{\rm r}$ can be used in combination with the pre-outburst r-band magnitude ($r=19.95\pm0.04$) and $A_r$ extinction to calculate the distance. $A_r$ can be derived from the reddening to the source $E(B-V)$, using $A_r/E(B-V)=2.271$ \citep{Schlafly2011} and $A_V=3.1\, E(B-V)$ \citep{Savage1979,Fitzpatrick2004}. We apply several methods to constrain the colour excess $E(B-V)$ and/or distance:

\begin{itemize}
    \item The equivalent width (EW) of the interstellar line \ion{K}{i} 7699 is a known tracer of the reddening \citep{Munari1997}, and covered during E5. Nearby telluric features limited the precision of its measurement as they affected the determination  of the adjacent continuum. We employed a Monte Carlo approach to simulate $10^{5}$ spectra using as a seed the original data and its observational uncertainties, and obtained $\rm{EW}=0.21\pm 0.07\,{\rm \AA}$, implying $E(B-V)=0.9\pm 0.3$, which results in $d=2.0\pm 0.7 \, {\rm kpc}$.
    \item The DIB at 8621 $\rm{\AA}$ is also correlated with reddening \citep{Wallerstein2007}. Its equivalent width of $\rm{EW}=0.18\pm 0.06\,{\rm \AA}$, measured with the same method as above, corresponds to $E(B-V)=0.8\pm 0.3$ and to a distance of $d=2.2\pm 0.7 \, {\rm kpc}$. 
    \item The hydrogen column density was estimated from X-ray fitting to be in the range of $N_H=(2.26-4.1)\cdot 10^{21} \rm cm^{-2} $ \citep{Oconnor2023, Draghis2023}, with a mean value of $N_H=(3.2\pm 0.9)\cdot 10^{21} \rm cm^{-2} $, from which we derive $E(B-V)= 0.47\pm 0.13$ \citep{Guver2009} and $d=3.0\pm 0.5\, {\rm kpc}$.
\end{itemize}
\citet{Megier2009} studied the correlation of the interstellar Ca II doublet (H and K) with the distance to early-type stars beyond $1\,{\rm kpc}$, taking into account the line saturation. J1727 appears to fulfil the conditions to apply this correlation [EW(K)/EW(H) > 1.3; and being no further than a few hundred parsecs from the Galactic plane]. This results in $d=3.2\pm 0.6\,  \rm{kpc}$.

We calculated the conditional probability resulting from the above methods. This results in $E(B-V)=0.57 \pm 0.11$, as well as $d=2.7 \pm 0.3\, {\rm kpc}$, and therefore, a height over the Galactic plane of $z=0.48 \pm 0.05\, {\rm kpc}$. 

\section{Conclusions}

We present multi-epoch optical spectroscopy during the outburst of the newly discovered BHT J1727. We cover the main stages of the outburst, and report on the detection of diverse outflow features, including blue-shifted absorptions sometimes accompanied by red-shifted excesses (i.e. P-Cygni profiles), broad emission wings, and flat-top profiles, in the H Balmer series. We also report the detection of similar features in the higher excitation \ion{He}{ii}-$4686\,\rm{\AA}$ line. The presence of red-shifted absorptions accompanied by blue emission wings in a number of epochs suggests the presence of inflows. Based on the current information available, we propose the following configuration for this system: a BHT with an early K-type companion, $P_{\rm orb}\sim 7.6\, {\rm h,}$ and placed at a distance of $d=2.7 \pm 0.3\, {\rm kpc}$ ($z=0.48 \pm 0.05\, {\rm kpc}$ over the Galactic plane).

\begin{acknowledgements}
DMS, TMD, and MAP acknowledge support by the Spanish Ministry of Science via the Plan de Generacion de conocimiento: PID2020-120323GB-I00 and PID2021-124879NB-I00; as well as a Europa Excelencia grant (EUR2021-122010). We thank Tom Marsh for the use of \textsc{molly} software. We are thankful to the GTC staff for their prompt and efficient response at triggering the time-of-opportunity program at the source of the spectroscopy presented in this Letter. Based on observations collected at the European Southern Observatory under ESO programme 105.20LK.002. This work has made use of data from the European Space Agency (ESA) mission {\it Gaia} (\url{https://www.cosmos.esa.int/gaia}), processed by the {\it Gaia} Data Processing and Analysis Consortium (DPAC, \url{https://www.cosmos.esa.int/web/gaia/dpac/consortium}). Funding for the DPAC has been provided by national institutions, in particular the institutions participating in the {\it Gaia} Multilateral Agreement. \textsc{pyraf} is the python implementation of \textsc{iraf} maintained by the community.
\end{acknowledgements}

%
%

\bibliographystyle{aa} 
\bibliography{bibliography} 

\begin{appendix}
\section{Compilation of possible outflow features.}

\begin{table*}
\caption{Compilation of possible outflow features.}
\begin{tabular}{ccccccccccc}
Epoch  & H$\delta$ & H$\gamma$ & \ion{He}{ii}-$4686\,\rm{\AA}$ & H$\beta$ & \ion{He}{i}-$5876\,\rm{\AA}$ &  H$\alpha$ & ML\\
 & & & & & & &   \\
\hline \\
E1 &blue-abs &blue-abs &blue-abs &blue-abs & $-$ &blue-abs &  d/o\\
E2 &blue-abs & $-$ & $-$ & $-$ & $-$& broad wings & d/d\\
E3 &blue-abs &blue-abs & skewed &blue-abs & $-$ & P-Cyg & o/ba\\
E4 & broad-abs &  iP-Cyg & red-abs & broad-abs & flat-top & broad-abs &o/d\\
E5 &blue-abs &blue-abs &blue-abs &blue-abs & $-$ & blue-abs & o/o \\
E6 &blue-abs &blue-abs & $-$ &blue-abs & $-$ &blue-abs & o/o\\
E7 &iP-Cyg &  iP-Cyg/flat-top & red-abs &  iP-Cyg & $-$ & broad wings & ba/o\\
E8 &blue-abs &blue-abs &blue-abs &blue-abs/flat-top & $-$ &blue-abs & o/o\\
E9 &red-abs &  red-abs & $-$ &  skewed & $-$ & skewed & d/d\\
E10 &broad-abs & broad-abs & $-$ &  broad-abs/skewed & skewed & broad-abs & o/ba\\
E11 &blue-abs &blue-abs & blue-abs &blue-abs/flat-top & blue-abs/flat-top &blue-abs & o/o\\
E12 &broad-abs & $-$ & $-$ & flat-top & $-$&  $-$ & d/d\\
E13 &broad-abs & blue-abs & $-$ & blue-abs & $-$&  blue-abs & d/d\\
E14 &broad-abs & blue-abs & flat-top & blue-abs & $-$&  blue-abs & o/o\\
E15 &broad-abs & $-$ & $-$ & red-abs & $-$&  broad-abs &  d/d\\
E16 &blue-abs &blue-abs & $-$ & blue-abs/flat-top & $-$&  blue-abs& o/o\\
E17 &broad-abs & $-$  & $-$ & $-$ & skewed &  skewed &  d/ba\\
E18 & broad-abs  & blue-abs  & flat-top & blue-abs & blue-abs &  skewed & d/ba\\
E19 & broad-abs  & red-abs  & $-$  & red-abs & $-$  &   red-abs & o/ba\\
E20 & broad-abs  & skewed  & flat-top  & blue-abs & $-$  &   $-$ & d/d\\
\end{tabular}
\tablefoot{Compilation of visually identified features deviating from a pure disc emission, including blue-shifted absorptions (blue-abs), red-shifted absorptions (red-abs), broad absorptions (broad-abs), flat top profiles (flat-top), skewed emission profiles (skewed), P-Cygni-like (P-Cyg) inverse P-Cygni-like (iP-Cyg), broad emission wings (broad wings), and red/blue emission excess tails (red/blue tail). ML refers to the machine learning classification of the $\rm{H{\gamma}}$ and $\rm{H{\delta}}$  profiles: disc (d), outflows (o) (in the form of blue-shifted absorption, broad wings, or P-Cygni), and broad absorption component (ba).}
\label{tab:outflows}
\end{table*}

\end{appendix}

\end{document}